\documentclass[review]{elsarticle}

\usepackage{lineno,hyperref}
\modulolinenumbers[5]

\journal{Journal of \LaTeX\ Templates}

\usepackage{amsmath,amssymb,amsfonts,bm, xcolor}
\begin{document}

\begin{frontmatter}

\title{Spin-wave mediated interactions for Majority Computation using Skyrmions and Spin-torque Nano-oscillators}
\tnotetext[t1]{This work was supported in part by Competitive Research Programme (CRP) under Grant NRF-CRP12-2013-01, in part by Singapore Ministry of Education (MOE) Grant MOE2017-T2-1-114, in part by MOE Academic Research Fund Tier 1, and in part by the NUS Start-up Grant.}
\author{Venkata Pavan Kumar Miriyala\corref{cor1}}%
\ead{elevpkm@nus.edu.sg}

\author{Zhifeng Zhu}
\author{Gengchiau Liang}

\author{Xuanyao Fong\corref{cor2}}%
\address{Department of Electrical and Computer Engineering, \\
                National University of Singapore, Singapore, 117583.}
\ead{kelvin.xy.fong@nus.edu.sg}
\cortext[correspondingauthor]{Corresponding authors: Venkata Pavan Kumar Miriyala and Xuanyao Fong}
\begin{abstract}
Recent progress in all-electrical nucleation, detection and manipulation of magnetic skyrmions has unlocked the tremendous potential of skyrmion-based spintronic devices. Here, we show via micromagnetic simulations that the stable magnetic oscillations of STNO radiate spin waves (SWs) that can be scattered in the presence of skyrmions in the near vicinity. Interference between SWs emitted by the STNO and SWs scattered by the skyrmion gives rise to interesting dynamics that leads to amplification or attenuation of STNO's magnetic oscillations. In the presence of strong \textcolor{red}{Dzyaloshinskii-Moriya interaction} (DMI), the amplified magnetic oscillations evolve into a new skyrmion. These interactions between skyrmions and STNOs are found to be identical for both Neel-type and Bloch-type skyrmions, and are not observed between domain walls and STNOs. These findings offer a novel perspective in processing information using single skyrmions and we propose a 3-bit majority gate for logic applications.

\end{abstract}


\end{frontmatter}

\section{\label{sec:level1} Introduction}
Skyrmions are topologically protected spin textures originating from the competing interactions such as the exchange and Dzyaloshinskii-Moriya interactions (DMI) \cite{Dzya58, Moriya, Boulle2016,Luchaire2016}. They are highly stable and can avoid getting pinned to structural defects or impurities due to their topological stability \cite{Nagaosa2013, Sampaio2013}. Furthermore, depending on the material properties and the strength of DMI, skyrmions can be as small as 5 nm \cite{Nagaosa2013} enabling ultra-dense information storage and processing devices. Recent studies also highlight the possibilities of efficient nucleation, annihilation and current-driven motion of skyrmions using current densities as low as 100 A-cm$^{-2}$ \cite{Sampaio2013,Romming636, Jonietz1648}. Subsequently, skyrmions are envisaged as promising candidates for novel memory \cite{Muller, Yang}, logic \cite{Xing2016, Diana2018, pinna18, kang2016}, oscillators \cite{Grac16, Jin18}, and artificial neuron \cite{chenk, Azam} devices.

In addition, it is observed that skyrmions interact with the spin waves (SWs) in different modes: breathing modes \cite{Kim2014}, azimuthal modes \cite{Mruc2018, Masa2012}, and gyrotropic modes \cite{Masa2012, Mruc2017}. As compared to \cite{Kim2014, Mruc2018, Masa2012, Mruc2017}, here we demonstrate a different mechanism to achieve electrical manipulation or interaction with skyrmions using SWs. Our proposed mechanism is based on our observation in micromagnetics that SWs get elastically scattered by a skyrmion, which then appears to be a source of radiating SWs. The effect may be understood by considering skyrmions as rigid particles due to topological protection, and they forbid topologically trivial SWs to pass through them \cite{garst,Iwasaki}. Earlier work has shown that this scattering mechanism is observed for planar SWs \cite{garst}. In this work, we utilize a spin-torque nano-oscillator (STNO) \cite{chen2016, KIM2012} as a SW source, and show in micromagnetics that the dynamics of the STNO can be altered due to the scattering of SWs by skyrmions. Furthermore, if the DMI is sufficiently strong in the magnetic system, the SW-mediated interaction between the STNO and a skyrmion can lead to the nucleation of another skyrmion. Therefore, our results presented here offer a novel perspective in electrical manipulation and interaction with skyrmions, and we demonstrate how these interactions between STNOs and skyrmions can be leveraged to implement a majority logic gate.

The rest of the paper is organized as follows. The simulation methodology used in this study is described in Section.~\ref{sec:level2}. The SW mediated interactions between STNOs and skyrmions are introduced and discussed in Section.~\ref{sec:level3}. In Section.~\ref{sec:level4}, we demonstrate the implementation of a skyrmion and STNO-based 3-bit majority logic gate. Finally, Section.~\ref{sec:level5} concludes this paper. 

\section{\label{sec:level2} Simulation Methodology}
The open-source MuMax3 micromagnetic tool \cite{Arne2014} is used to study the interactions between STNOs and skyrmions. The orientation of the free layer magnetization, $\bm{M}$, is modelled as the unit vector $\bm{\vec{m}}$. The time evolution of $\bm{\vec{m}}$ is given by the Landau-Lifshitz-Gilbert-Slonczewski \cite{Landau1935, Gilbert2004} equation written as 

\begin{eqnarray}
\left(1+\alpha^{2}\right)\frac{d\bm{\vec{m}}}{dt} & = & -\gamma\bm{\vec{m}}\times\left(\bm{\vec{B}_\text{eff}} + \alpha \bm{\vec{m}}\times\bm{\vec{B}_\text{eff}}\right)\nonumber\\
&& - \beta(\epsilon-\alpha\epsilon')\bm{\vec{m}}\times\left(\bm{\vec{m}}\times\bm{\vec{m}_\text{p}}\right)\nonumber\\
&& + \beta(\epsilon'-\alpha\epsilon)\bm{\vec{m}}\times\bm{\vec{m}_\text{p}} \\
\beta & = & \frac{\hbar\gamma J_\text{z}}{e{t_\text{FL}}M_\text{S}}\\ 
\epsilon & = & \frac{P\cdot\Lambda^{2}}{\left(\Lambda^{2}+1\right)+\left(\Lambda^{2}-1\right)\left(\bm{\vec{m}}\cdot\bm{\vec{m}_\text{p}}\right)}
\end{eqnarray}

where $\gamma$ = 1.76085$\times$$10^{11}$ rad/s$\cdot$T is the gyromagnetic ratio, $\alpha$ is the damping parameter, $\bm{\vec{B}_\text{eff}}$ is the total effective magnetic field of the free layer, $J_\text{z}$ is the current density along the $\it{z}$-axis, $\textit{e}$ is the electron charge, $M_\text{S}$ is the saturation magnetization, $t_\text{FL}$ is the free layer thickness, $\bm{\vec{m}_\text{p}}$ is the pin layer magnetization, $\textit{P}$ is the spin polarization, $\Lambda$ is the Slonczewski parameter that characterizes the spacer layer, and $\epsilon'$ is the secondary spin-torque parameter.

The first term on the right-hand side of (1), models the dynamics of $\bm{\vec{m}}$ induced by the $\bm{\vec{B}_\text{eff}}$, which consists of magnetostatic field, Heisenberg exchange field, Dzyaloshinskii-Moriya (DM) exchange field, magneto-crystalline anisotropy field, and the externally applied magnetic field. The competition between non-collinear DM exchange field and the other field components of $\bm{\vec{B}_\text{eff}}$ leads to the formation of skyrmions in the free layer magnetization. The remaining two terms on the right-hand side of (1), models the dynamics of $\bm{\vec{m}}$ due to the current-induced spin-transfer torque (STT) \cite{Xiao2004, PavanTED, Pavansispad}. 

In all the simulations performed in this work, the simulation volume is discretized into rectangular meshes with 2 nm$\times$2 nm$\times$0.4 nm cell sizes. The material parameters considered are typical of Co/Pt bilayer systems with perpendicular magnetic anisotropy (PMA) \cite{Sampaio2013}: $M_\text{S}$ =  5.8$\times$$10^{5}$ A/m; exchange stiffness constant, $A_\text{ex}$ = 1.5$\times$$10^{-11}$ J/m; interfacial DMI constant, D = 3 mJ/m$^{2}$; PMA constant, $K_\text{u}$ = 0.8$\times$$10^{6}$ J/m$^{3}$; $t_\text{FL}$ = 0.4 nm. The $\bm{\vec{m}_\text{p}}$, $\alpha$, $\textit{P}$, $\Lambda$, and $\epsilon'$ are assumed to be [0.1736, 0, 0.9848], 0.03, 0.4, 1.2, and $10^{-3}$, respectively.

\section{\label{sec:level3} Results \& Discussion}

The device structure (Fig.~\ref{fig:fig_1} (a-b)) used in our simulation study consists of a 1 $\mu$m diameter circular multi-layer ferromagnetic system with two distinct circular regions: the skyrmion region and the nano-contact STNO region. Both the circular regions have a diameter of 40 nm and are each connected to a current sources denoted as $I_\text{1}$ and $I_\text{2}$, respectively. The distance between the centers of the two circular regions is denoted as \textbf{\textit{d}}. The dimensions of the ferromagnetic system were chosen to minimize the influence of edge effects on the simulation results. In the proposed device, we first inject the current $I_\text{1}$ with current density of 7.25$\times10^{11}$ A/m$^{2}$ and pulse width ($\tau_\text{1}$) of 5 ns into the skyrmion region (i.e. green colored circular region in Fig.~\ref{fig:fig_1} (a)). $I_\text{1}$ exerts the STT on the local magnetization and if the torque is sufficiently large, the local magnetization is switched from one state to the other and finally stabilizes into a skyrmion \cite{Romming636,Ma2015} due to the DMI interactions. Next, $I_\text{1}$ is reduced to zero and the overall magnetization is allowed to relax for 2 ns. After relaxation, the current $I_\text{2}$ with current density of \textcolor{red}{5.4985$\times10^{11}$ A/m$^{2}$} and pulse width ($\tau_\text{2}$) of 5 ns is injected into the STNO region (red colored circular region in Fig.~\ref{fig:fig_1} (a)). $I_\text{2}$ exerts the STT on the local magnetization and balances out the damping-like torque generated by the $\bm{\vec{B}_\text{eff}}$. As a result, the local magnetization, $\bm{\vec{m}}$ oscillates continuously as shown by the in-plane component of $\bm{\vec{m}}$ plotted in Fig.~\ref{fig:fig_1} (c) \cite{chen2016, KIM2012}. The frequency of these oscillations is estimated to be \textcolor{red}{63 GHz} (Fig.~\ref{fig:fig_1} (d)). These continuous magnetic oscillations radiate SWs into the sample (Fig.~\ref{fig:fig_1} (e)). \textcolor{red}{The average wavelength ($\bm{\lambda}$) of these SWs is found to be 100 nm approximately using the spatial fast Fourier transforms to the data obtained in micromagnetics.}

Let us now investigate the interactions between skyrmions and STNOs. First, in the absence of skyrmion, the local magnetization in the STNO region continuously oscillates and radiates SWs into the sample (Fig.~\ref{fig:fig_2} (a-e)). Interestingly in the presence of a skyrmion at \textbf{\textit{d}} = 100 nm, the magnitude of magnetic oscillations are amplified (Fig.~\ref{fig:fig_2} (f-j)). As shown in Fig.~\ref{fig:fig_2} (j), the STNO becomes unstable resulting in significant disturbances in ferromagnet's magnetization at $\tau_\text{2}$ = 4 ns. However, if the skyrmion is at \textbf{\textit{d}} = 125 nm, the magnitude of magnetic oscillations are attenuated (Fig.~\ref{fig:fig_2} (k-o)). Similar amplification and attenuation in STNO's magnetic oscillations are observed when the skyrmion is at \textbf{\textit{d}} = 150 nm (Fig.~\ref{fig:fig_2} (p-t)) and 175 nm (Fig.~\ref{fig:fig_2} (u-y)), respectively. This \textbf{\textit{d}}-dependent periodic trend in the amplification and attenuation of STNO's magnetic oscillations can be due to the interference between SWs emitted by the STNO ($\it{SWs}$$_\text{emitted}$) and SWs scattered by the skyrmion ($\it{SWs}$$_\text{scattered}$). Note that the skyrmions studied here are of Neel-type due to interfacial nature of DMI interactions in Co/Pt bilayer systems \cite{Finocchio}. The same type of interaction between SWs and the skyrmion is also observed in case of Bloch-type skyrmions. Bloch-type skyrmions commonly exist in non centro-symmetric or cubic magnets with bulk-DMI interactions \cite{Finocchio}. \textcolor{red}{Although it is not physical, here we created the Bloch-type skyrmion by fictitiously modeling the nature of DMI interactions as bulk instead of interfacial.} The dynamics of STNOs under the influence of Neel-type skyrmion and Bloch-type skyrmion are shown in Fig.~\ref{fig:fig_4} (a-e) and (f-j), respectively. The amplification in STNO's magnetic oscillations is observed in both the cases and therefore, we may conclude that this interaction between skyrmions and STNOs is identical for both Neel-type and Bloch-type skyrmions. 

Moreover, this effect is observed only in the presence of skyrmions but not in the presence of domain walls \cite{midd}. As shown in Fig.~\ref{fig:fig_4} (k-o) and (p-t), a Neel-type domain wall does not scatter the SWs and thus, the STNO steadily oscillates without amplification or attenuation regardless of whether a Neel-type domain wall is present. Similar results are observed in case of Bloch-type domain walls as well (not shown here). Therefore, we may conclude that the interaction between skyrmions and STNOs we have presented thus far is unique for the skyrmions owing to their topological properties. Furthermore, the results shown so far are obtained by placing the skyrmion on the right-hand side of STNO at \textbf{\textit{d}}. We observed similar results even if the skyrmion is placed on left-hand side or on top or bottom-side of STNO (not shown here) and that edge effects or the magnetization orientation along the edges do not play any role in these interactions between STNOs and skyrmions. 

Skyrmions behave as rigid particles and do not allow $\it{SWs}$$_\text{emitted}$ to pass through their magnetic texture. Therefore, $\it{SWs}$$_\text{emitted}$ gets scatterred back by the skyrmion \cite{garst,Iwasaki}. For example, the magnitude of $\it{SWs}$$_\text{scattered}$ when the skyrmion is located at \textbf{\textit{d}} = 100 nm is shown in Fig.~\ref{fig:fig_3} (c). The $\it{SWs}$$_\text{emitted}$ getting scattered from all around the skyrmion appearing as if the skyrmion is reradiating the SWs. However, the magnitude of $\it{SWs}$$_\text{scattered}$ is strong in particular directions (Fig.~\ref{fig:fig_3} (c)) as SWs experience the Magnus force resulting from the topological magnetic texture of skyrmions \cite{garst,Iwasaki}. Despite the differences in magnitude, the $\it{SWs}$$_\text{scattered}$ has same wavelength as compared to the $\it{SWs}$$_\text{emitted}$. In addition, no phase difference is observed between $\it{SWs}$$_\text{emitted}$ and $\it{SWs}$$_\text{scattered}$. Constructive interference between $\it{SWs}$$_\text{emitted}$ and $\it{SWs}$$_\text{scattered}$ can amplify the STNO's magnetic oscillations, whereas destructive interference attenuates the STNO's magnetic oscillations. The equations describing $\it{SWs}$$_\text{emitted}$ (Eq.~\ref{eq:inter1}) and $\it{SWs}$$_\text{scattered}$ (Eq.~\ref{eq:inter2}), as observed from the results obtained from the micromagnetics can be written as
\begin{eqnarray}
\label{eq:inter1}
\it{SWs}_\text{emitted} & \propto & A e^{\frac{k}{k_\text{f}}x-\omega t} cos(kx -\omega t)\\
\label{eq:inter2}
\it{SWs}_\text{scattered} & \propto & B e^{\frac{k}{k_\text{f}}x-\omega t - \frac{\Phi}{2\Phi_\text{f}}} cos(kx -\omega t-\Phi), \Phi = 2k(\textbf{\textit{d}}-R)\\
\label{eq:inter3}
\it{SWs}_\text{total} & = & \it{SWs}_\text{emitted} + \it{SWs}_\text{scattered}
\end{eqnarray}
where, A and B are constants that govern the wave amplitudes,  $\omega$ is the angular frequency, $\it{k}$ = 2$\pi$/$\bm{\lambda}$ is the wave vector, $k_\text{f}$ and $\Phi_\text{f}$ are the fitting parameters used to reproduce the micromagnetics results, $\Phi$ is the phase difference, and $\it{R}$ is the radius of the skyrmion. To simplify the study, the Magnus force acting on the SWs has been neglected. When the skyrmion is at \textbf{\textit{d}}, $\it{SWs}$$_\text{emitted}$ travels from $\it{x}$ = 0 (i.e. the centre of STNO region) to \textbf{\textit{d}}-$\it{R}$ (i.e. the boundary of skyrmion's magnetization) and gets scattered by the skyrmion. Thus, the $\it{SWs}$$_\text{scattered}$ starts propagating back from $\it{x}$ = \textbf{\textit{d}}-$\it{R}$ to 0 to reach the STNO. Subsequently, at $\it{x}$ = 0, $\it{SWs}$$_\text{scattered}$ accumulates a total phase difference of 2$\it{k}$(\textbf{\textit{d}}-$\it{R}$) as compared to $\it{SWs}$$_\text{emitted}$. \textcolor{red}{The zero reflection phase difference observed between $\it{SWs}$$_\text{emitted}$ and $\it{SWs}$$_\text{scattered}$ can be due to different modes of interaction between skyrmions and SWs. As skyrmion's magnetic texture changes during the SW interaction \cite{Kim2014, Mruc2018, Masa2012, Mruc2017}, skyrmion's boundary can be treated as soft. Generally, when waves encounter soft boundaries, no phase difference arises between incident and reflected waves. For instance, when one end of the string is open-ended (fixed), the reflected wave will have 0$^\circ$ (180$^\circ$) phase change \cite{jones1986,WinNT,PSU01}.}

From the equations Eq.~\ref{eq:inter1} to Eq.~\ref{eq:inter3}, we find that $\it{SWs}$$_\text{emitted}$ and $\it{SWs}$$_\text{scattered}$ add constructively for \textbf{\textit{d}} = $\bm{\lambda}$+ n$\bm{\lambda}$/4 whereas they add destructively for \textbf{\textit{d}} = $\bm{\lambda}$+ m$\bm{\lambda}$/4, where n and m are positive even and odd integers, respectively. In other words, for every $\bm{\lambda}$/4 change in \textbf{\textit{d}}, the STNO's magnetization dynamics change between amplification and attenuation regimes. These results are in agreement with the micromagnetic simulation results. For instance, when $\bm{\lambda}$ = 100 nm, and \textbf{\textit{d}} = 100 nm (i.e. n = 0) and 150 nm (i.e. n = 2), the oscillations of STNOs are amplified (See Fig.~\ref{fig:fig_2} (f-j) and (p-t)). Similarly, when $\bm{\lambda}$ = 100 nm, and \textbf{\textit{d}} = 125 nm (i.e. m = 1) and 175 nm (i.e. m = 3), the oscillations of STNO are significantly attenuated (See Fig.~\ref{fig:fig_2} (k-o) and (u-y)). 

Next, to further verify the analytical theory, we obtained the trends in amplification and attenuation of STNO's oscillations using different $\bm{\lambda}$ and \textbf{\textit{d}} through micromagnetic simulations. \textcolor{red}{To obtain different $\bm{\lambda}$, we added an external magnetic field ($\bm{\vec{B}_\text{ext}}$) and used different magnitudes of $I_\text{2}$ in the STNO region. As per the SW dispersion relation Eq.~\ref{eq:inter5} \cite{Jamali2016}, applying $\bm{\vec{B}_\text{ext}}$ modulates the frequency of STNO's magnetic oscillations ($\it{f}$). Depending on $\it{f}$, the velocity at which the SWs spread out in the sample varies giving rise to SWs with different $\bm{\lambda}$. 
\begin{eqnarray}
\label{eq:inter5}
\it{f} & = & \frac{\gamma}{2 \pi}\sqrt{(\bm{\vec{B}_\text{eff}} + \bm{\vec{B}_\text{ext}})(\bm{\vec{B}_\text{eff}} + \bm{\vec{B}_\text{ext}} + \mu_{0}M_\text{S}) + \frac{(\mu_{0}M_\text{S})^2}{4}(1-e^{-2kt_\text{FL}})}
\end{eqnarray}}
As expected, when $\bm{\lambda}$ = 120 nm, 80 nm and 40 nm, the STNOs alternate between amplification and attenuation regimes for every $\bm{\lambda}$/4 change in \textbf{\textit{d}} i.e. 30 nm, 20 nm, and 10 nm, respectively (See Table~\ref{tab:table_1}). If $\bm{\lambda}$ = 120 nm, the amplification in STNO oscillation is observed when \textbf{\textit{d}} = 120 nm (i.e. n = 0) and 180 nm (i.e. n = 2). If $\bm{\lambda}$ = 80 nm, amplification in STNO oscillation is observed when \textbf{\textit{d}} = 120 nm (i.e. n = 2) and 160 nm (i.e. n = 4). If $\bm{\lambda}$ = 40 nm, amplification in STNO oscillation is observed when \textbf{\textit{d}} = 120 nm, 140 nm, 160 nm, and 180 nm (i.e. for n = 8, 10, 12, and 14, respectively). Similarly, for all the different $\bm{\lambda}$ studied, the attenuation of magnetic oscillations is observed when m = 1, 3, 5, ...etc. Therefore, we can conclude that this interaction between STNOs and skyrmions arise due to the interference between the SWs emitted by the STNOs and the SWs scattered by the skyrmions. 
\begin{table*}[t]
\centering
\begin{tabular}{|c|c|c|c|c|c|c|c|c|c|}
\hline
$\textbf{\textit{d}}$ (nm) & 100 & 110 & 120 & 130 & 140 & 150 & 160 & 170 & 180 \\
\hline
$\textbf{\textit{$\bm{\lambda}$}}$ = 120 nm & C & C & C & D & D & D & C & C & C\\
\hline
$\textbf{\textit{$\bm{\lambda}$}}$ = 80 nm   & D & C & C & D & D & C & C & D & D\\
\hline
$\textbf{\textit{$\bm{\lambda}$}}$ = 40 nm   & C & D & C & D & C & D & C & D & C\\
\hline
\end{tabular}
\caption{Wavelength ($\textbf{\textit{$\bm{\lambda}$}}$) and distance (\textbf{\textit{d}}) dependent STNO's magnetic oscillations amplification due to constructive interference (C) and attenuation due to destructive interference (D)}
\label{tab:table_1}
\end{table*}

Interestingly, if the DMI is sufficiently strong (i.e. 3.5 mJ/m$^{2}$) in the sample, it is observed that the amplified magnetic oscillations results in nucleation of a new skyrmion. As shown in Fig.~\ref{fig:fig_5} (a-e) and (f-j), a new skyrmion forms in the STNO region when the first skyrmion is at \textbf{\textit{d}} = 100 nm and 150 nm, respectively. This is due to the DMI interactions that transform significant distortions in the ferromagnet's magnetization into skyrmions \cite{Liu2015, Giordano2016}. Note that with increase in DMI constant from 3 mJ/m$^{2}$ to 3.5 mJ/m$^{2}$, the skyrmion's radius, $\it{R}$ increases from 45 nm to 50 nm \cite{Fert2017}. Therefore, SW mediated interactions between STNOs and skyrmions can be used to manipulate the presence or absence of skyrmions, opening up a new perspective in information processing using skyrmions. As an example, we next demonstrate a 3-bit STNO-based majority gate for logic applications.

\section{\label{sec:level4} Skyrmion and STNO-based 3-bit Majority gate}
The device structure assumed here consists of a of a 1 $\mu$m diameter circular multi-layer ferromagnetic system having four distinct circular regions with 40 nm diameters: three input bit regions (A, B and C) and one output bit region (O) (See Fig.~\ref{fig:fig_6}). The absence or presence of skyrmion in the input bit regions represents bit 0 or 1, respectively. To perform 3-bit majority computation on the three input bits, a STNO is placed in the output bit region. The distance between the centers of output bit region and each input bit region is described as \textbf{\textit{d}}. As discussed in Section.~\ref{sec:level3}, when \textbf{\textit{d}} = $\bm{\lambda}$+ n$\bm{\lambda}$/4, SWs emitted by the STNO and SWs scattered by the skyrmion add constructively. As a result, a new skyrmion forms in the STNO region representing output 1. 

\begin{table*}[t]
\centering
\begin{tabular}{|c|c|c|c|}
\hline
A & B & C & O  \\
\hline
0 & 0 & 0 & 0 \\
\hline
0 & 0 & 1 & 0 \\
\hline
0 & 1 & 0 & 0 \\
\hline
0 & 1 & 1 & 1 \\
\hline
1 & 0 & 0 & 0 \\
\hline
1 & 0 & 1 & 1 \\
\hline
1 & 1 & 0 & 1 \\
\hline
1 & 1 & 1 & 1 \\
\hline
\end{tabular}
\caption{Truth Table for 3-bit Majority Gate}
\label{tab:table_2}
\end{table*}

\textcolor{red}{For this device, we assumed \textbf{\textit{d}} to be 200 nm, i.e. n = 4, where $\bm{\lambda}$ is 100 nm. First, the input bits are written into the specified regions using the charge currents with current density of 7.25$\times10^{11}$ A/m$^{2}$ and pulse width ($\tau_\text{1}$) of 5 ns. Next, we induce continuous magnetic oscillations in the STNO region using charge current with current density of 5.4985$\times10^{11}$ A/m$^{2}$ and pulse width ($\tau_\text{2}$) of 8 ns. If all the three input bits are in state 0, the STNO steadily oscillates and no skyrmion forms in the STNO region, i.e. output 0 (See Fig.~\ref{fig:fig_6} (a-d)). If one of the three input bits is in state 1, the STNO oscillations get amplified and a new skyrmion forms in the STNO region in 8 ns of time delay (See Fig.~\ref{fig:fig_6} (e-h)), i.e output 1. If any two of the input bits are 1, the STNO's magnetic oscillations are further amplified and the amount of amplification is roughly doubled due to the presence of two skyrmions. Consequently, a new skyrmion nucleates with in the pulse width of 6 ns, i.e. output 1 (See Fig.~\ref{fig:fig_6} (i-l)). Similarly, if all three input bits are 1, the STNO's magnetic oscillations get strongly amplified and a new skyrmion nucleates in the output region with in the pulse width of 4 ns, i.e. output 1 (See Fig.~\ref{fig:fig_6} (m-p)). The presence of output skyrmion in the STNO region can be electrically read using the non-collinear magnetoresistance (NCMR) effect \cite{Hanneken2015,Crum2015, Kube}. The output skyrmion in the STNO region must be annihilated prior to the next logic operation and this annihilation can be performed by passing the current with current density of and pulse width through the STNO region. In addition, by fixing the input bit A to 0, a 2-bit AND gate can be realized and by fixing the input bit A to 1, a 2-bit OR gate can be realized (See Table.~\ref{tab:table_2}). Note that in all the logic gates proposed, skyrmions in the STNO regions must be annihilated prior to logic operation.}

\textcolor{red}{Finally, a 5-bit and 7-bit majority gates can be designed by placing the input bits at a radial distance of $\bm{\lambda}$+ n$\bm{\lambda}$/4 from the center of output region. In 5-bit majority gate where 5 input bits are placed at a radius of 200 nm from the center of output bit region, $\tau_\text{2}$ must be less than the time required to generate the output skyrmion when 2 of 5 input bits are state 1, i.e. 6 ns. Similarly, $\tau_\text{2}$ in case of 7-bit majority gate must be less than the time required to generate the output skyrmion when 3 of 7 input bits are in state 1, i.e. 4 ns.}

\section{\label{sec:level5} Conclusion}
In conclusion, we showed that skyrmions strongly interfere with the dynamics of spin torque nano-oscillators (STNOs). The stable magnetic oscillations of STNOs are shown to be amplified or attenuated depending on the position of skyrmion in the near vicinity. This interaction between topologically non-trivial skyrmions and STNOs is found to be unique and does not exist between domain walls and STNOs. We have also shown that this interaction between STNOs and skyrmions arises from the constructive or destructive interference between spin waves (SWs) emitted by STNOs and the SWs scattered by skyrmions. Furthermore, we showed that if Dzyaloshinskii-Moriya interaction (DMI) is sufficiently strong in the sample, the amplified magnetic oscillations cause a new skyrmion to be nucleated. Finally, we demonstrated the skyrmion and STNO-based majority gate device. 

\section*{References}

\bibliography{main}
\begin{figure}[!hbtp]
\centering
\includegraphics[width=4.5in]{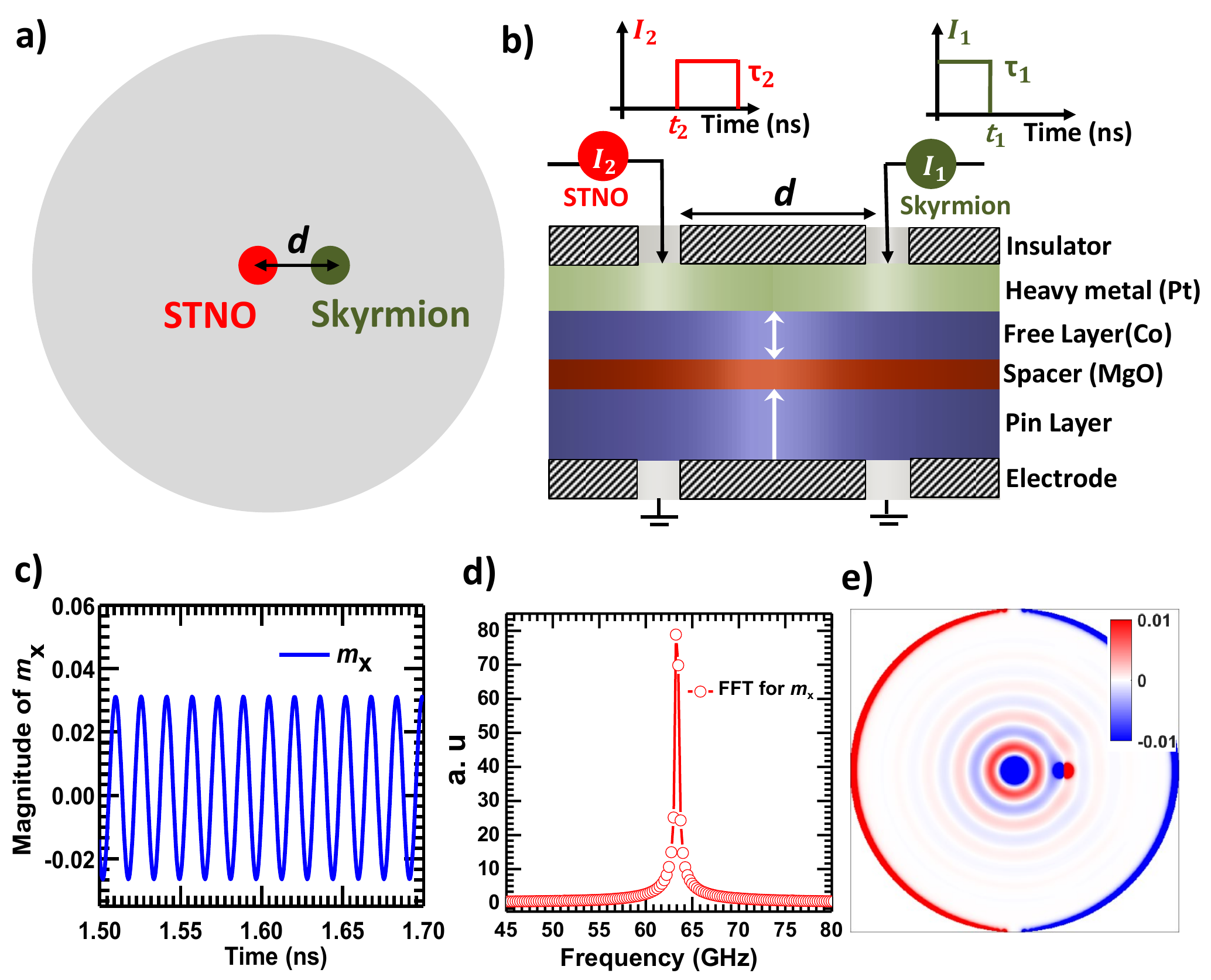}
\caption{(a) The top view of the proposed device structure with two regions marked-the spin torque nano-oscillator (STNO) region (red circle), and the skyrmion region (green circle). \textbf{\textit{d}} is the distance between the centers of STNO and skyrmion regions. (b) The cross-sectional view of the device consisting of a multi-layer perpendicular ferromagnetic system with the charge current sources at STNO and skyrmion regions. First, the charge current ($I_\text{1}$) flowing through the skyrmion region generates skyrmion. Then, the charge current ($I_\text{2}$) flowing through the STNO region makes the magnetization continuously precess. As the magnetization in STNO region precess; (c) shows the oscillating in-plane magnetization component ($m_\text{x}$) in the STNO region, (d) shows the frequency spectrum of the STNO's magnetic oscillations, and (e) shows the propagation of spin-waves from the STNO region. Blue and red dumbbell like structure in (e) indicates the skyrmion's magnetization component. The inset in (e) the scaled color map used for plotting $m_\text{x}$.}
\label{fig:fig_1}
\addtocounter{figure}{0}
\end{figure}
\begin{figure*}[!htbp]
\centering
\includegraphics[width=4.5in]{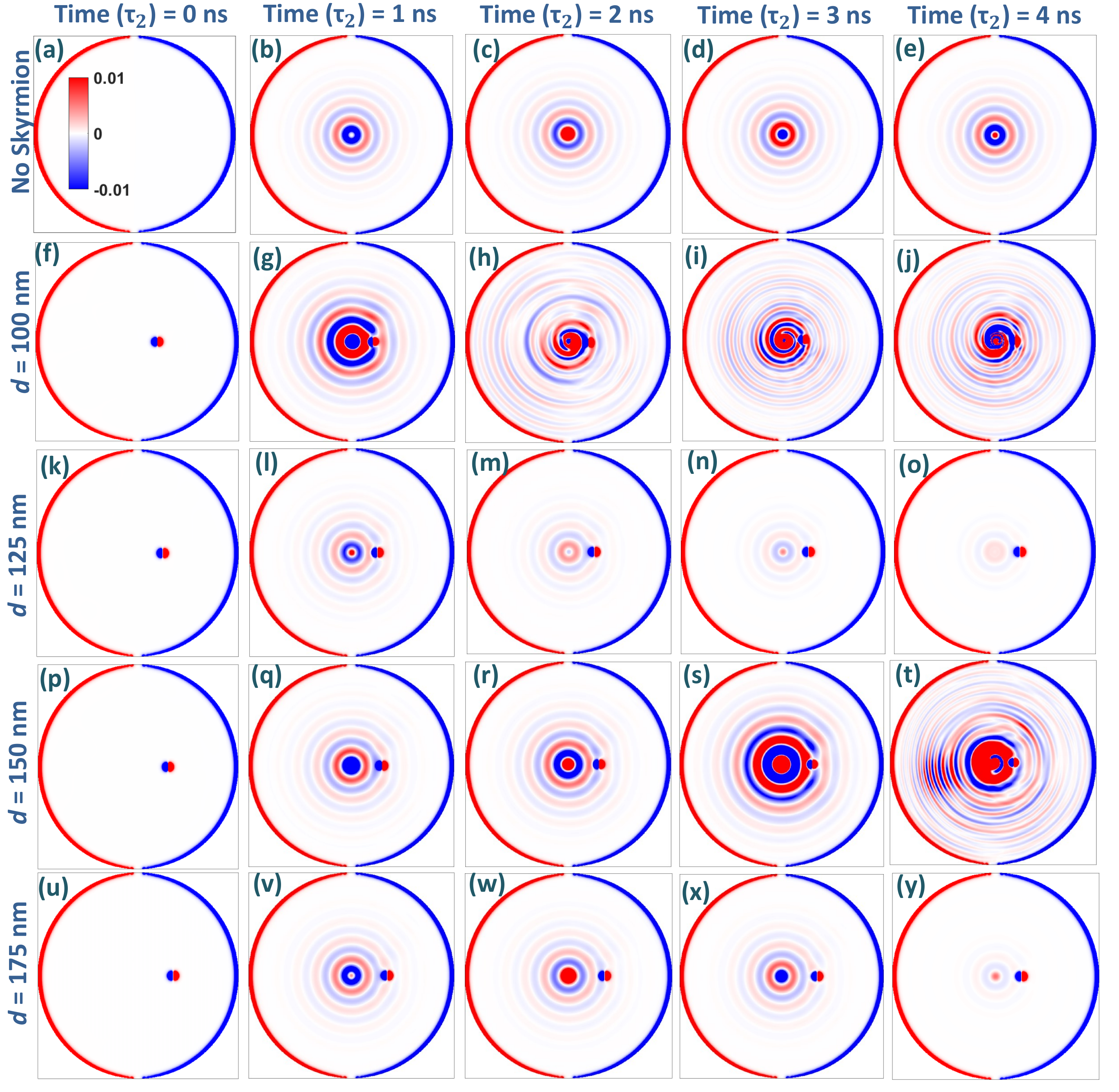}
\caption{The magnitude of ferromagnet's magnetization along the $\it{x}$-direction ($m_\text{x}$). (a-e) in the absence of skyrmion, where the STNO’s magnetization steadily oscillates. (f-j) and (p-t) in the presence of skyrmion at \textbf{\textit{d}} = 100 nm and 150 nm, respectively, where STNO's magnetic oscillations are amplified. (k-o) and (u-y) in the presence of skyrmion at \textbf{\textit{d}} = 125 nm and 175 nm, respectively, where the STNO's magnetic oscillations are attenuated. Wavelength ($\bm{\lambda}$) of the SWs in this case is observed to be 100 nm, approximately. The inset in (a) shows the color map used for plotting $m_\text{x}$.}
\label{fig:fig_2}
\addtocounter{figure}{0}
\end{figure*}
\begin{figure*}[!hbtp]
\centering
\includegraphics[width=4.5in]{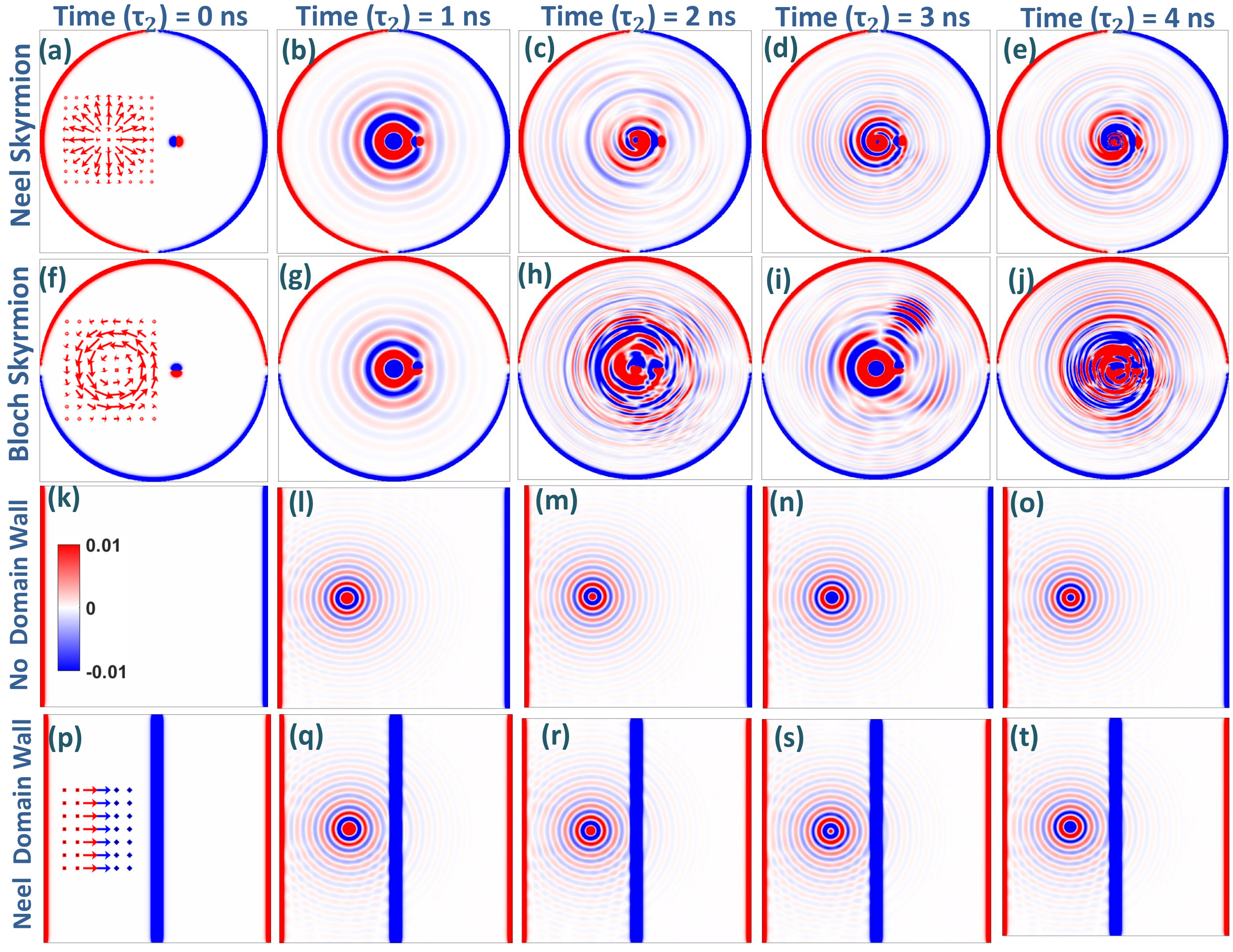}
\caption{The magnitude of ferromagnet's magnetization along the $\it{x}$-direction ($m_\text{x}$). (a-e) shows the formation of a new Neel skyrmion when the old Neel skyrmion is located at \textbf{\textit{d}} = 100 nm. (f-j) shows the formation of a new Bloch skyrmion when the old Bloch skyrmion is located at \textbf{\textit{d}} = 100 nm. (k-o) shows the SWs in the absence of a domain wall in the rectangular ferromagnetic system, and (p-t) shows the SWs in the presence of a domain wall. In contrast to the skyrmion cases, the STNO in (p-t) remains stable and the magnitude of SWs remains constant. The inset in (k) shows the color map used for plotting $m_\text{x}$.}
\label{fig:fig_4}
\addtocounter{figure}{0}
\end{figure*}
\begin{figure*}[t]
\centering
\includegraphics[width=4.5in]{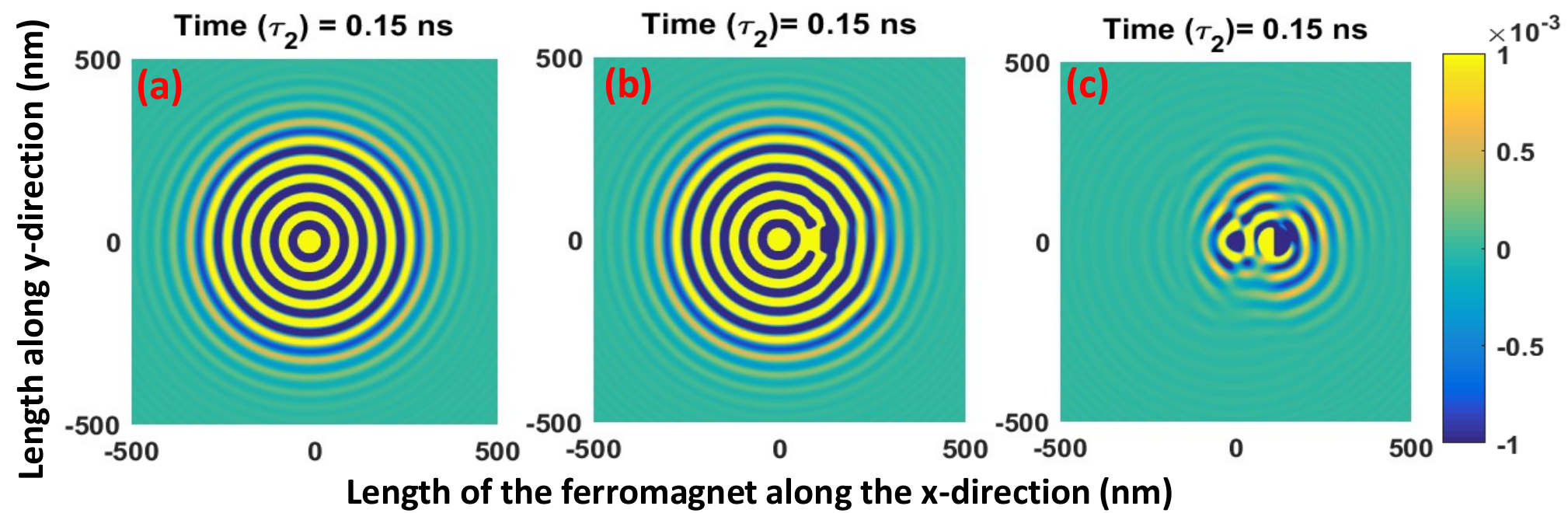}
\caption{The magnitude of ferromagnet's magnetization along the $\it{x}$-direction ($m_\text{x}$). (a) in the absence of skyrmion, (b) in the presence of skyrmion at {\textbf{\textit{d}}} = 100 nm, and (c) SWs scattered by the skyrmion. Note that (c) is obtained by calculating the difference between magnitude of magnetization in the absence (a) and presence (b) of skyrmion. Therefore, (c) also visualizes an additional wave component in the STNO region that results from the amplification of STNO's magnetic oscillations.}
\label{fig:fig_3}
\addtocounter{figure}{0}
\end{figure*}
\begin{figure*}[!hbtp]
\centering
\includegraphics[width=4.5in]{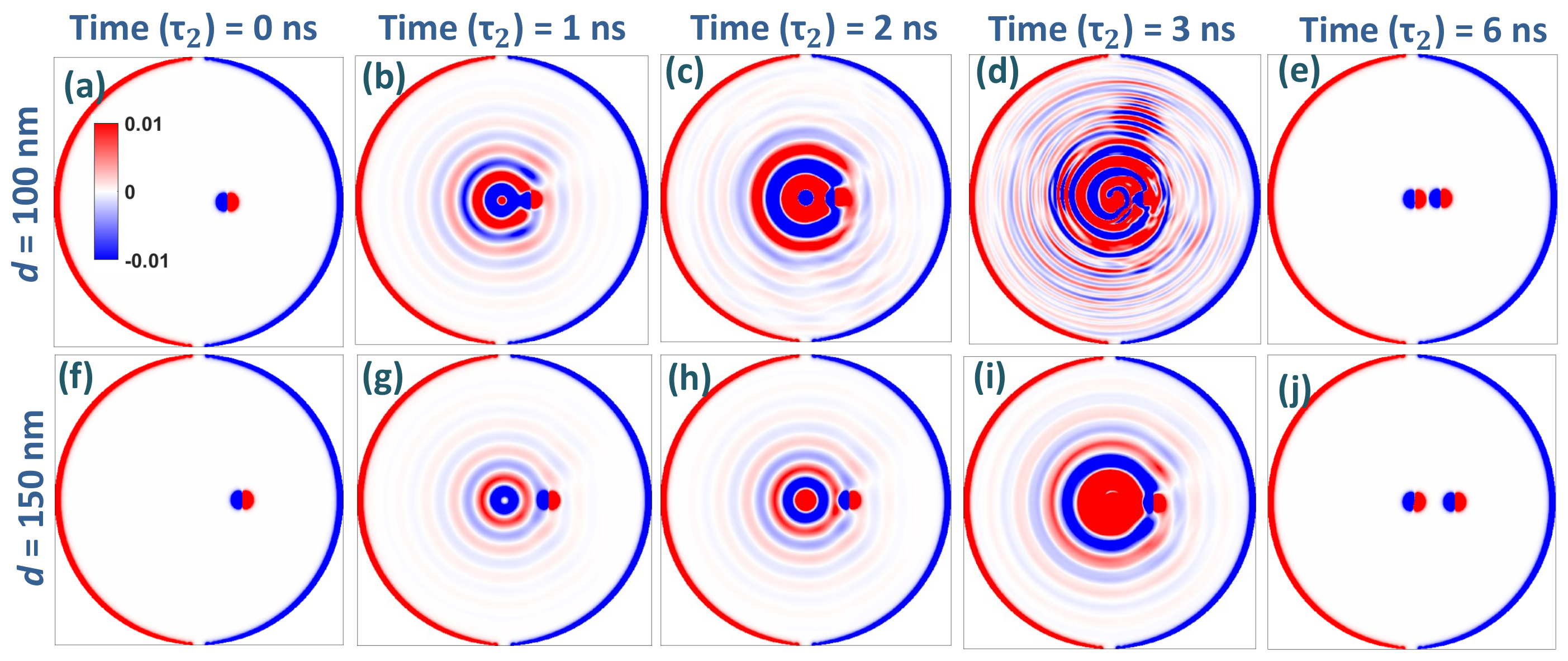}
\caption{The magnitude of ferromagnet's magnetization along the $\it{x}$-direction ($m_\text{x}$) when the DMI is increased from 3 mJ/m$^{2}$ to 3.5 mJ/m$^{2}$. (a-e), and (f-j) shows the nucleation of a new skyrmion when the old skyrmion is located at \textbf{\textit{d}} = 100 nm, and 150 nm, respectively. Due to strong DMI, large disturbances in the ferromagnet's magnetization lead to the nucleation of a new skyrmion. The inset in (a) shows the color map used for plotting $m_\text{x}$.}
\label{fig:fig_5}
\addtocounter{figure}{0}
\end{figure*}
\begin{figure*}[!hbtp]
\centering
\includegraphics[width=4.5in]{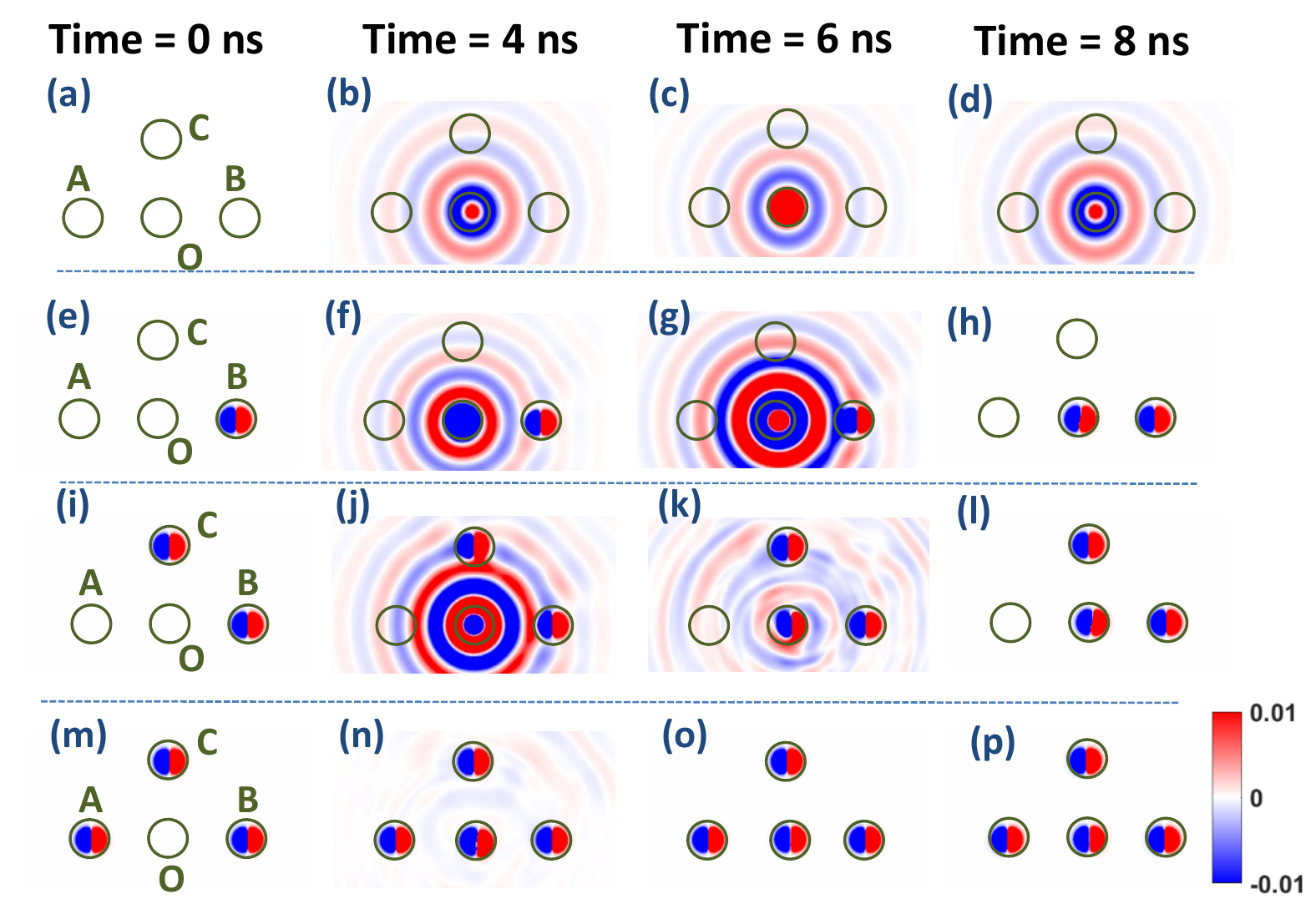}
\caption{The magnitude of ferromagnet's magnetization along the $\it{x}$-direction ($m_\text{x}$) in the proposed majority gate device. Bits 0 and 1 are represented by the absence and presence of skyrmion in three input bit regions (A, B and C) and one output bit region (O). The time evolution of the majority gate (a-d) when all three inputs are in state 0, (e-h) when one of the three inputs is state 1, (i-l) when two of the three inputs are in state 1 and (m-p) when all the three inputs are state in 1. Other possible combinations of input bits shown in Table.~\ref{tab:table_2} are not shown here. The inset shows the color map used for plotting $m_\text{x}$.}
\label{fig:fig_6}
\addtocounter{figure}{0}
\end{figure*}

\end{document}